\documentclass[sn-mathphys,Numbered]{sn-jnl}

\usepackage{graphicx}%
\usepackage{multirow}%
\usepackage{amsmath,amssymb,amsfonts}%
\usepackage{amsthm}%
\usepackage{mathrsfs}%
\usepackage[title]{appendix}%
\usepackage{xcolor}%
\usepackage{textcomp}%
\usepackage{manyfoot}%
\usepackage{booktabs}%
\usepackage{algorithm}%
\usepackage{algorithmicx}%
\usepackage{algpseudocode}%
\usepackage{listings}%

\theoremstyle{thmstyleone}%
\theoremstyle{thmstyletwo}%

\theoremstyle{thmstylethree}%

\begin{document}

\title[Article Title]{Time Optimal Qubit Computer}


\author*[1]{\fnm{Peter} \sur{Morrison}}\email{peter.morrison@uts.edu.au}

\affil*[1]{\orgdiv{Department of Maths and Physical Sciences}, \orgname{University of Technology, Sydney}, \orgaddress{\street{Broadway}, \city{Sydney}, \postcode{2007}, \state{NSW}, \country{Australia}}}

\abstract{

We present a number of new physical systems that may be addressed using methods of time dependent transformation. A recap of results available for two-state systems is given, with particular emphasis on the AC stark effect. We give some results that are not well known, including the full solution for a two state system in a static electric field with arbitrary direction. Connection with established theorems in time optimal quantum control is given, and a full discussion outlines some advanced results in matrix calculus. In particular, we derive a set of matrix gates relevant to quantum information theory and computation using time optimal unitary operators, and define the hyperbolic equivalent of the quantum brachistochrone problem.
}




\maketitle

\section{Introduction}

The topic of time optimal quantum control is a rapidly expanding field;
as there are many new methodologies emerging from the sector, this
paper will seek to bring together a number of disparate approaches
and unify them under the principle of the quantum brachistochrone.
Time optimal quantum control and the science of the quantum brachistochrone,
originally developed by Carlini et. al. \cite{carlini2005quantum, carlini2006time, carlini2007time} is an elegant
formalism that seeks to find paths of least time on different complex
projective spaces. In this paper we will extend this formulation of
quantum mechanics to new territory; we shall discuss the various transformations
that can be applied to the most basic time optimal quantum systems,
and demonstrate how this relates physical systems with quite different
observed behaviour. Other works by the author have focused on the
problem of qutrit optimal control \cite{morrison2019time2}, time optimal
control of a relativistic electron \cite{morrison2019time}, and indeed
have examined in passing the problem of transformations of a time
optimal qubit \cite{morrison2012time}. It is this last problem that this
paper shall focus on, and we shall show how the quantum brachistochrone
principle implies that all paths of least time can be related by transformation
to a certain fundamental solution. We shall analyse how this system
of analysis can be extended from spherical spaces, such as is given
by the SU(2)/Bloch sphere correspondence, to more exotic hyperbolic
spaces where time is taken to be a complex number. 

\section{AC/DC Stark Effect}

We shall briefly discuss the AC and DC Stark effects, as these are
the systems which are closely aligned with the more complicated systems
that we shall examine in this paper. For the DC Stark effect, the
Hamiltonian matrix is constant, and defined by the matrix operator:

\begin{equation}
\tilde{H}=\left[\begin{array}{cc}
E+\Delta & Ve^{-i\phi}\\
Ve^{i\phi} & E-\Delta
\end{array}\right]
\end{equation}
\begin{equation}
=E\mathbf{1}+\mathbf{n}.\mathbf{\sigma}
\end{equation}
In the diagonal reference frame, we have eigenvalues $\lambda=E\pm\Omega=E\pm\sqrt{\Delta^{2}+V^{2}}$.
We have eigenvectors $\left|\Omega_{+}\right\rangle $,$\left|\Omega_{-}\right\rangle $,
which we may write as the solution matrix: 
\begin{equation}
\hat{W}=\dfrac{1}{\sqrt{2}}\left[\begin{array}{cc}
-\dfrac{Ve^{-i\phi}}{\Delta-\Omega} & -\dfrac{Ve^{-i\phi}}{\Delta+\Omega}\\
1 & 1
\end{array}\right]
\end{equation}
with $\Omega^{2}=\Delta^{2}+V^{2}$. This has inverse given by:
\begin{equation}
\hat{W}^{-1}=\dfrac{1}{\sqrt{2}\Omega}\left[\begin{array}{cc}
-Ve^{i\phi} & -(\Delta-\Omega)\\
+Ve^{i\phi} & \Delta+\Omega
\end{array}\right]
\end{equation}
and hence we may diagonalise the Hamiltonian via 
\begin{equation}
\hat{W}^{-1}\tilde{H}\hat{W}=\left[\begin{array}{cc}
E+\Omega & 0\\
0 & E-\Omega
\end{array}\right]=\tilde{L}
\end{equation}
As in this case $\Omega$ and $E$ are not explicitly time dependent,
we may evaluate the time evolution operator using the telescopic property
of the above transformations. 
\begin{equation}
\hat{U}(t,0)=\exp\left(-i\int_{0}^{t}\tilde{H}(s)ds\right)=\exp\left(-i\tilde{H}t\right)
\end{equation}
\begin{equation}
=\exp\left(-it.\hat{W}\tilde{L}\hat{W}^{-1}\right)
\end{equation}
\begin{equation}
=\hat{W}\exp\left(-it.\tilde{L}\right)\hat{W}^{-1}
\end{equation}
Using the previous formulae, we obtain the compact form of the unitary
as:
\begin{equation}
\hat{U}(t,0)=\left[\begin{array}{cc}
\cos\Omega t-\dfrac{i\Delta\sin\Omega t}{\Omega} & -\dfrac{iVe^{-i\phi}\sin\Omega t}{\Omega}\\
\dfrac{iVe^{+i\phi}\sin\Omega t}{\Omega} & \cos\Omega t+\dfrac{i\Delta\sin\Omega t}{\Omega}
\end{array}\right]
\end{equation}
which has the required properties $\hat{U}(t,0)\hat{U}^{\dagger}(t,0)=\hat{U}^{\dagger}(t,0)\hat{U}(t,0)=\mathbf{1}$.
This closes this problem, as this operator is sufficient to specify
all physical behaviour in the system. 

We shall now move to the AC Stark effect, where it is still possible
to diagonalise the Hamiltonian matrix, but the eigenvalues are periodic
and time dependent. This is a much more interesting situation, as
the solution is not known in closed form. Consider a modified version
of the previous problem; by making the off-diagonal elements in the
Hamiltonian matrix oscillatory, we obtain a new physical system. This
is well understood in terms of the rotating wave approximation, but
let us not make this adjustment to begin with. The Hamiltonian operator
we shall use may be written as:

\begin{equation}
\tilde{H}=\left[\begin{array}{cc}
+E & Ve^{-i\phi}\cos\omega t\\
Ve^{i\phi}\cos\omega t & -E
\end{array}\right]
\end{equation}
This has time dependent eigenvalues $\lambda_{\pm}=\pm\sqrt{E^{2}+V^{2}\cos^{2}\omega t}=\pm\Omega(t)$,
and we may write the solution matrix and inverse in the form: 
\begin{equation}
\hat{W}(t)=\dfrac{1}{\sqrt{2}}\left[\begin{array}{cc}
-\dfrac{Ve^{-i\phi}\cos\omega t}{E-\Omega} & -\dfrac{Ve^{-i\phi}\cos\omega t}{E+\Omega}\\
1 & 1
\end{array}\right]
\end{equation}
\begin{equation}
\hat{W}^{-1}(t)=\dfrac{1}{\sqrt{2}\Omega}\left[\begin{array}{cc}
-Ve^{i\phi}\cos\omega t & -(E-\Omega)\\
+Ve^{i\phi}\cos\omega t & E+\Omega
\end{array}\right]
\end{equation}
Performing the diagonalisation procedure as before, we find $\hat{W}^{-1}(t)\tilde{H}(t)\hat{W}(t)=\Omega(t)\left[\begin{array}{cc}
1 & 0\\
0 & -1
\end{array}\right]=\tilde{L}(t)$. Note that this essentially complicates the exponentiation of the
Hamiltonian matrix, as the diagonal matrix is now explicitly dependent
on time. We are unable to perform the technique in the previous section
and must change our approach. The standard approach using the rotating
wave approximation is obviously valid, but as we are looking for exact
solutions, this is a theoretical dead end. We must modify the way
in which we view this problem in order to find a way through.

\section{Quantum Brachistochrone}

We shall discuss a simple derivation of the quantum brachistochrone
equation. Our approach differs from Carlini et. al. in that we shall
show that the simplest way to find this equation is through standard
techniques from the von Neumann equation, of which this is a special
case. Regardless, the action principle for the quantum brachistochrone
is defined by:

\begin{equation}
S=\int1dt+\int\lambda_{1}[\mathrm{Tr}(\tilde{H}\tilde{F})]dt+\int\lambda_{1}[\mathrm{Tr}\left(\dfrac{\tilde{H}^{2}}{2}\right)-k]dt
\end{equation}
A small note on this formulation; we take the expression $\Delta E=\sqrt{\left\langle \Psi\right|\left.\tilde{H}^{2}(t)\left|\Psi\right\rangle -\left(\left\langle \Psi\right|\tilde{H}(t)\left|\Psi\right\rangle \right)^{2}\right.}$,
being the energy dispersion or variance of the operator $\tilde{H}(t)$
with respect to the state $\left|\Psi\right\rangle $. In this sense,
the factor $1$ in the first integral, which represents the condition
of least time, is then given by 
\begin{equation}
1=\dfrac{\sqrt{\left\langle \Psi\right|\left.\tilde{H}^{2}(t)\left|\Psi\right\rangle -\left(\left\langle \Psi\right|\tilde{H}(t)\left|\Psi\right\rangle \right)^{2}\right.}}{\Delta E}
\end{equation}
To obtain the quantum brachistochrone, with some work, one can follow
the calculations in \cite{carlini2005quantum, morrison2008time}. The basic
idea is to use the standard form of the Euler-Lagrange equations,
applied to the different variables $\left|\Psi\right\rangle ,\tilde{H}(t),\lambda_{1},\lambda_{2}$.
We shall not go into this here, but refer the interested reader to
the works contained in \cite{carlini2005quantum, carlini2006time}. The method we
shall take is much more simple and direct. The von Neumann equation
may be written as:
\begin{equation}
i\dfrac{d\hat{A}}{dt}=[\tilde{H}(t),\hat{A}]
\end{equation}
for any Hermitian operator $\hat{A}(t)$ evolving unitarily under
the influence of the system Hamiltonian $\tilde{H}(t)$. In terms
of our time optimal system, we choose the operator which is given
by $\hat{A}(t)=\tilde{H}(t)+\tilde{F}(t)$, which is Hermitian as
we have assumed both the system Hamiltonian and associated constraint
may be specified by Hermitian matrices. In this case, we very easily
arrive at the set of coupled differential equations:
\begin{equation}
i\dfrac{d}{dt}(\tilde{H}(t)+\tilde{F}(t))=[\tilde{H}(t),\tilde{F}(t)]
\end{equation}
which defines the quantum brachistochrone, originally appearing in
\cite{carlini2005quantum, carlini2006time} and expounded upon in \cite{morrison2008time}. The constraint
laws we associate to this expression are easily derived from the Euler-Lagrange
equations for $\lambda_{1},\lambda_{2}$; we have immediately that
$\mathrm{Tr}(\tilde{H}\tilde{F})=0$, also $\mathrm{Tr}\left(\dfrac{\tilde{H}^{2}}{2}\right)-k=0$.
The second of these we term the isotropic condition. These expressions-
quantum brachistochrone \& constraints- define a unique Hamiltonian
matrix which evolves the state of the system from one configuration
to another in least time. We shall now show how one may apply this
method to some simple quantum systems.

\section{Time Optimal Hamiltonian}

We shall now solve the quantum brachistochrone equation for a simple
example which will form the basis for the rest of the calculations
in this paper. This method of solving the quantum brachistochrone
is explored in \cite{morrison2012time, morrison2019time, morrison2019time2} and is somewhat different to that
used in \cite{carlini2005quantum, carlini2006time, carlini2007time}. What we shall do is find a clever eigenvalue
decomposition which renders the equations of motion into a form which
is readily solved. To this end, let us choose a Hamiltonian matrix
via $\tilde{H}(t)=\lambda_{x}(t)\hat{\sigma}_{x}+\lambda_{y}(t)\hat{\sigma}_{y}$,
the associated constraint satisfying $\mathrm{Tr}(\tilde{H}\tilde{F})=0$
is then given by $\tilde{F}=\Omega(t)\hat{\sigma}_{z}$. Evaluating
the matrix operators, we have:

\begin{equation}
\tilde{H}(t)=\left[\begin{array}{cc}
0 & \epsilon(t)\\
\epsilon^{*}(t) & 0
\end{array}\right]
\end{equation}
where $\epsilon(t)=\lambda_{x}(t)-i\lambda_{y}(t)$, and the constraint
\begin{equation}
\tilde{F}(t)=\left[\begin{array}{cc}
\Omega(t) & 0\\
0 & -\Omega(t)
\end{array}\right]
\end{equation}
hence the quantum brachistochrone equations read as the coupled set
of DEs specified by the matrix equation:
\begin{equation}
i\dfrac{d}{dt}\left[\begin{array}{cc}
\Omega(t) & \epsilon(t)\\
\epsilon^{*}(t) & -\Omega(t)
\end{array}\right]=2\Omega\left[\begin{array}{cc}
0 & -\epsilon(t)\\
\epsilon^{*}(t) & 0
\end{array}\right]
\end{equation}
from which we conclude that $\Omega(t)=\Omega$, i.e. the constraint
is a constant matrix, and hence the optimal Hamiltonian may be written
as:
\begin{equation}
\tilde{H}_{opt}(t)=R\left[\begin{array}{cc}
0 & e^{2i\omega t}\\
e^{-2i\omega t} & 0
\end{array}\right]
\end{equation}
with the particular choice of initial conditions $\epsilon(0)=\epsilon^{*}(0)=R$,
and we have taken $\omega=\Omega$. The reason for this will become
apparent, as we are interested in some simple changes to the Hamiltonian
and constraint it makes sense to use the parameter $\omega$ in place
of $\Omega$ to avoid confusion. We shall now construct the time evolution
operator for this system. To accomplish this we require access to
the matrix of eigenvectors, which we term the 'eigenmatrix' or 'fundamental
matrix'. Writing this down, it is simple to see that this matrix is
given by:
\begin{equation}
\hat{W}(t)=\dfrac{1}{\sqrt{2}}\left[\begin{array}{cc}
e^{2i\omega t} & -e^{2i\omega t}\\
1 & 1
\end{array}\right]
\end{equation}
In terms of the Hamiltonian matrix, the unitary operator is given
by:
\begin{equation}
\hat{U}(t,s)=\exp\left(-i\int_{s}^{t}\tilde{H}(\tau)d\tau\right)
\end{equation}
We also have the time evolution equation for the Hamiltonian, being
$\hat{U}(t,s)\tilde{H}(s)\hat{U}^{\dagger}(t,s)=\tilde{H}(t)$. Let
us apply the eigenmatrix to the Hamiltonian; we find:
\begin{equation}
\hat{W}^{-1}(t)\tilde{H}(t)\hat{W}(t)=\hat{L}=R\left[\begin{array}{cc}
1 & 0\\
0 & -1
\end{array}\right]
\end{equation}
hence we may write $\tilde{H}(t)=\hat{W}(t)\hat{L}\hat{W}^{\dagger}(t)$,
applying this to separate times $s$ and $t$ we therefore have $\hat{U}(t,s)=\hat{W}(t)\hat{W}^{-1}(s)$
in order to satisfy the time evolution equation. By observation it
is simple to see that we have $\hat{W}^{-1}(t)=\hat{W}^{\dagger}(t)$,
therefore we may write:
\begin{equation}
\hat{U}(t,s)=\hat{W}(t)\hat{W}^{\dagger}(s)=e^{i\phi}\left[\begin{array}{cc}
e^{i\phi} & 0\\
0 & e^{-i\phi}
\end{array}\right]
\end{equation}
where $\phi=\omega(t-s)$. This is the first of the unitary operators
we shall explore in this paper. All the other unitary operators we
shall consider will be transformations of this operator in one way
or another. It is a simple exercise in linear algebra to show that
we satisfy the quantum brachistochrone, time evolution equation and
basic unitarity as well as time translation invariance. To do so,
we must show that the expressions $\hat{U}_{Q}(t,s)\hat{U}_{Q}^{\dagger}(t,s)=\hat{U}_{Q}^{\dagger}(t,s)\hat{U}_{Q}(t,s)=\mathbf{1}$,
to demonstrate unitarity; that $\hat{U}_{Q}(t,s)\tilde{H}_{Q}(s)\hat{U}_{Q}^{\dagger}(t,s)=\tilde{H}_{Q}(t)$
to show that the action of the unitary on the Hamiltonian shifts it
forward in time; and that $\hat{U}_{Q}(t_{1},s)\hat{U}_{Q}(s,t_{2})=\hat{U}_{Q}(t_{1},t_{2})$
to illustrate time translation invariance. All of these properties
are easily established using elementary linear algebra so we will
not look into them further.

\section{Brachistochrones via Unitary Transforms}

We shall now look at some interesting results from time optimal quantum
control, which relate to the subgroups we have relied on in our proof.
Firstly, we shall take as axioms the basic equations of time optimal
quantum control, being:

\begin{equation}
i\dfrac{d}{dt}(\tilde{H}+\tilde{F})=[\tilde{H},\tilde{F}]=\tilde{H}\tilde{F}-\tilde{F}\tilde{H}
\end{equation}

\begin{equation}
\mathrm{Tr}(\tilde{H}\tilde{F})=0
\end{equation}
and
\begin{equation}
\mathrm{Tr}\left(\dfrac{\tilde{H}^{2}}{2}\right)=k<\infty
\end{equation}
Assume for now that we have some constant, invertible matrix which
we call $\hat{A}$. The above equations are then invariant under the
isomorphic transformation $\hat{X}\rightarrow\hat{A}\hat{X}\hat{A}^{-1}$
where $\hat{X}$ is either $\tilde{H}$ (Hamiltonian) or $\tilde{F}$
(constraint). Writing $\hat{X}_{A}=\hat{A}\hat{X}\hat{A}^{-1}$ with
$\hat{X}_{1}=\hat{X}$, $\mathbf{1}$ being the identity transformation,
$\hat{X}$ standing as a placeholder for either the constraint, Hamiltonian
or unitary operator it is simple to see that the equations of time
optimal quantum control will then change to:
\begin{equation}
i\hat{A}\dfrac{d}{dt}(\tilde{H}+\tilde{F})\hat{A}^{-1}=\hat{A}\tilde{H}\tilde{F}\hat{A}^{-1}-\hat{A}\tilde{F}\tilde{H}\hat{A}^{-1}
\end{equation}
Noting that, on the proviso that $\hat{a}$ is some constant matrix
which does not change with time, we are free to move it inside the
time differential, also using $\hat{A}\hat{A}^{-1}=\mathbf{1}$ we
may insert a factor onto the right hand side, resulting in the equation
\begin{equation}
i\dfrac{d}{dt}(\tilde{H}_{A}+\tilde{F}_{A})=[\tilde{H}_{A},\tilde{F}_{A}]
\end{equation}
The constraint laws may be similarly manipulated, using the cyclic
property of the trace we obtain variously $\mathrm{Tr}\left(\dfrac{\tilde{H}_{A}^{2}}{2}\right)=k$
and $\mathrm{Tr}(\tilde{H}_{A}\tilde{F}_{A})=0$, thereby demonstrating
the invariance of these equations under constant linear isomorphism.
Consider now the basic solution to the quantum brachistochrone equation
on SU(2), covered first in \cite{carlini2005quantum} and expanded upon in
various works by the author \cite{morrison2008time, morrison2012time, morrison2019time, morrison2019time2}. In the most
simple of situations, we have a constraint law given by $\tilde{F}=\Omega(t)\hat{\sigma}_{z}$,
following the analysis of \cite{morrison2012time} the optimal Hamiltonian
which drives the system from a chosen start point to terminal value
in least time is then derived by satisfying the quantum brachistochrone
equation, which we recognise as the von Neumann equation. In any case,
the solution to the equation is 
\begin{equation}
\tilde{H}(t)=R\left[\begin{array}{cc}
0 & e^{2i\omega t}\\
e^{-2i\omega t} & 0
\end{array}\right]
\end{equation}
with a constant constraint specified by $\Omega(t)=\Omega$, i.e.
$\tilde{F}=\Omega\hat{\sigma}_{z}$. We have shown in the previous
section why this is the relevant brachistochrone for this particular
quantum system. Now, consider that the unitary operator for this system
may be readily evaluated using eigenvalue decomposition, yielding
\begin{equation}
\hat{U}(t,s)=e^{i\phi}\left[\begin{array}{cc}
e^{i\phi} & 0\\
0 & e^{-i\phi}
\end{array}\right]
\end{equation}
with $\phi=\omega(t-s)$. We note that this is essentially the same
as one of the generators of SU(2), with an additional extra phase
term which does not contribute to any physical differences. Let us
now consider transformations of the unitary operator, in the same
way that we analysed isomorphic transformations of the Hamiltonian
operator. Assuming that the operator $\hat{T}$ is constant, it is
a simple exercise to show that 

\[
\hat{T}\hat{U}\hat{T}^{-1}=\hat{T}\exp\left(-i\int_{s}^{t}\tilde{H}(\tau)d\tau\right)\hat{T}^{-1}
\]

\begin{equation}
=\exp\left(-i\int_{s}^{t}\hat{T}\tilde{H}(\tau)\hat{T}^{-1}d\tau\right)
\end{equation}
Using the notation from before, we have $\hat{U}_{T}(t,s)=\hat{T}\hat{U}\hat{T}^{-1}$,
choosing the matrix $\hat{T}=\dfrac{1}{\sqrt{2}}\left[\begin{array}{cc}
1 & -i\\
-i & 1
\end{array}\right]$, we have transformed Hamiltonian:
\begin{equation}
\tilde{H}_{T}(t)=\hat{T}\tilde{H}(t)\hat{T}^{-1}=R\left[\begin{array}{cc}
-\sin2\omega t & \cos2\omega t\\
\cos2\omega t & \sin2\omega t
\end{array}\right]
\end{equation}
with a constraint that transforms to $\tilde{F}_{T}=-\Omega\left[\begin{array}{cc}
0 & -i\\
i & 0
\end{array}\right]=-\Omega\hat{\sigma}_{y}$. Computing the left hand side, we can therefore write:

\begin{equation}
\hat{U}_{T}(t,s)=\hat{T}\hat{U}\hat{T}^{-1}=e^{i\phi}\left[\begin{array}{cc}
\cos\phi & -\sin\phi\\
\sin\phi & \cos\phi
\end{array}\right]
\end{equation}
where, as before, $\phi=\omega(t-s)$. In this case, we can automatically
write down the exact solution for the unitary evolution under the
Hamiltonian operator for a quite complicated motion. Indeed, we have
the expression:

\[
\hat{U}_{T}(t,s)=\exp\left(-i\int_{s}^{t}\hat{T}\tilde{H}(\tau)\hat{T}^{-1}d\tau\right)
\]

\begin{equation}
=\exp\left(-i\int_{s}^{t}\left[\begin{array}{cc}
-\sin2\omega\tau & \cos2\omega\tau\\
\cos2\omega\tau & \sin2\omega\tau
\end{array}\right]d\tau\right)
\end{equation}

\begin{equation}
=e^{i\phi}\left[\begin{array}{cc}
\cos\phi & -\sin\phi\\
\sin\phi & \cos\phi
\end{array}\right]
\end{equation}
with $\phi=\omega(t-s)$, and appropriate modifications for the inclusion
of the parameter $R$. We notice that, up to a global phase ($e^{i\phi}$),
this is a rotation matrix which defines one of the subgroups of SU(2).
We shall now show that the operator $\hat{U}_{T}(t,s)$ has all the
required properties to be a well defined time evolution operator.
We note also that the problem of exponentiating the matrix on the
right-hand side of the expression is highly non-trivial, and not amenable
to analysis. However, we have managed through a clever trick to find
the answer. It is interesting that these subgroups arise naturally
as a result of this methodology. In a sense, SU(2) is too 'nice' to
be an accurate description of reality, as the components fit together
in a way which may not be reflective of the relative complexities
that arise in higher dimensional systems. 

To demonstrate that the matrix $\hat{U}_{T}(t,s)$ has the necessary
properties for a well-behaved time evolution operator, we must show
that it is unitary via $\hat{U}\hat{U}^{\dagger}=\hat{U}^{\dagger}\hat{U}=\mathbf{1}$,
and we must also demonstrate that it propagates the Hamiltonian forwards
in time via $\hat{U}(t,s)\tilde{H}(s)\hat{U}^{\dagger}(t,s)=\tilde{H}(t)$.
Let us carry out these basic tasks before calculating the remaining
subgroups for SU(2) using an identical method. The first property
is easily established, we have $\hat{U}_{T}(t,s)\hat{U}_{T}^{\dagger}(t,s)=\hat{U}_{T}^{\dagger}(t,s)\hat{U}_{T}(t,s)=\mathbf{1}$
by basic matrix multiplication. Again, computing the product, we have 

\[
\hat{U}_{T}(\phi)\tilde{H}_{T}(s)\hat{U}_{T}^{\dagger}(\phi)
\]

\begin{equation}
=R\left[\begin{array}{cc}
-\sin(2\omega s+2\phi) & \cos(2\omega s+2\phi)\\
\cos(2\omega s+2\phi) & \sin(2\omega s+2\phi)
\end{array}\right]
\end{equation}
\begin{equation}
=R\left[\begin{array}{cc}
-\sin(2\omega t) & \cos(2\omega t)\\
\cos(2\omega t) & \sin(2\omega t)
\end{array}\right]=\tilde{H}_{T}(t)
\end{equation}
where we have used sum of angles formulae from simple trigonometry.
We can see that this is indeed functioning correctly as a time evolution
operator should within this quantum space. We shall have recourse
to another operator, defined through the matrix $\hat{S}=\dfrac{1}{\sqrt{2}}\left[\begin{array}{cc}
i & -i\\
-1 & -1
\end{array}\right]$. The other operators defined by $\tilde{H}_{T^{-1}}(t)$, $\tilde{H}_{S}(t)$
and $\tilde{H}_{S^{-1}}(t)$ are readily shown to also obey the same
principles of unitarity and time translation for the Hamiltonian operator,
i.e. $\hat{U}_{Q}(\phi)\tilde{H}_{Q}(s)\hat{U}_{Q}^{\dagger}(\phi)=\tilde{H}_{Q}(t)$.
We shall write down the different Hamiltonian matrices and unitary
operators, we have:

\begin{equation}
\tilde{H}_{T}(t)=R\left[\begin{array}{cc}
-\sin(2\omega t) & \cos(2\omega t)\\
\cos(2\omega t) & \sin(2\omega t)
\end{array}\right]
\end{equation}
\begin{equation}
\hat{U}_{T}(\phi)=e^{i\phi}\left[\begin{array}{cc}
\cos\phi & -\sin\phi\\
\sin\phi & \cos\phi
\end{array}\right]
\end{equation}
as shown before;
\begin{equation}
\tilde{H}_{T^{-1}}(t)=R\left[\begin{array}{cc}
\sin(2\omega t) & \cos(2\omega t)\\
\cos(2\omega t) & -\sin(2\omega t)
\end{array}\right]
\end{equation}
\begin{equation}
\hat{U}_{T^{-1}}(\phi)=e^{i\phi}\left[\begin{array}{cc}
\cos\phi & \sin\phi\\
-\sin\phi & \cos\phi
\end{array}\right]
\end{equation}
where we note that we have $\hat{U}_{T^{-1}}(\phi)=\hat{U}_{T}^{*}(-\phi)$.
For the matrix $\hat{S}$:
\begin{equation}
\tilde{H}_{S}(t)=R\left[\begin{array}{cc}
-\cos(2\omega t) & \sin(2\omega t)\\
\sin(2\omega t) & \cos(2\omega t)
\end{array}\right]
\end{equation}
\begin{equation}
\hat{U}_{S}(\phi)=e^{i\phi}\left[\begin{array}{cc}
\cos\phi & \sin\phi\\
-\sin\phi & \cos\phi
\end{array}\right]
\end{equation}
and finally, for $\hat{S}^{-1}$ we obtain
\begin{equation}
\tilde{H}_{S^{-1}}(t)=R\left[\begin{array}{cc}
-\sin(2\omega t) & i\cos(2\omega t)\\
-i\cos(2\omega t) & \sin(2\omega t)
\end{array}\right]
\end{equation}
\begin{equation}
\hat{U}_{S^{-1}}(\phi)=e^{i\phi}\left[\begin{array}{cc}
\cos\phi & -i\sin\phi\\
-i\sin\phi & \cos\phi
\end{array}\right]
\end{equation}
where in all cases $\phi=\omega(t-s)$. Note that from the structure
of the results, we can see that there will be a relationship between
the matrices $\hat{T}$ and $\hat{S}$. We shall comment on this now.
A simple calculation shows that we have the matrix identities:
\begin{equation}
i\hat{S}^{-1}\hat{T}\hat{S}^{-1}=\hat{T}^{-1}
\end{equation}
and consequently we may also write
\begin{equation}
\hat{T}=-i\hat{S}\hat{T}^{-1}\hat{S}
\end{equation}
or in the alternative form $\hat{S}^{-1}\hat{T}=-i\hat{S}\hat{T}^{-1}$.
Using $\hat{S}^{-1}=\hat{S}^{\dagger}$ and likewise for $\hat{T}$,
this relation takes the form $\hat{S}^{\dagger}\hat{T}=-i\hat{S}\hat{T}^{\dagger}$,
which is a type of braiding relation. There is another matrix which
is closely related to these two; for this system it is given by the
Hadamard matrix which we define as:
\begin{equation}
\hat{V}=\dfrac{1}{\sqrt{2}}\left[\begin{array}{cc}
1 & 1\\
1 & -1
\end{array}\right]
\end{equation}
In this case, we have $\hat{V}^{\dagger}=\hat{V}^{-1}=\hat{V}$, so
the number of cases is reduced as the matrix is self-inverse and Hermitian.
We therefore only have to calculate $\tilde{H}_{V}(t)$ and $\hat{U}_{V}(t,s)$,
yielding:
\begin{equation}
\tilde{H}_{V}(t)=R\left[\begin{array}{cc}
\cos(2\omega t) & -i\sin(2\omega t)\\
i\sin(2\omega t) & -\cos(2\omega t)
\end{array}\right]
\end{equation}
\begin{equation}
\hat{U}_{V}(\phi)=e^{i\phi}\left[\begin{array}{cc}
\cos\phi & i\sin\phi\\
i\sin\phi & \cos\phi
\end{array}\right]=\hat{U}_{V}^{*}(-\phi)
\end{equation}
with $\hat{U}_{V}^{*}(\phi)=\hat{U}_{V}^{\dagger}(\phi)$. Under the
various different transformations, it is a simple process of matrix
multiplication to show that the constraint is changed to variously
\begin{equation}
\tilde{F}_{Q}=\left\{ \begin{array}{ccccc}
-\Omega\hat{\sigma}_{y}, & \Omega\hat{\sigma}_{y}, & \Omega\hat{\sigma}_{y}, & -\Omega\hat{\sigma}_{x}, & \Omega\hat{\sigma}_{x}\end{array}\right\} 
\end{equation}
for $\hat{Q}=\left\{ \hat{T},\hat{T}^{-1},\hat{S},\hat{S}^{-1},\hat{V}\right\} $.
Note the asymmetry in the constraint transformation, which matches
the asymmetry we observe in the unitary operators. There are other
transformations which we shall go into after a discussion of the matrix
of eigenvectors for the time optimal system.

\section{Transformations of Eigenmatrices}

Consider the quantum brachistochrone equation, we have shown in other
papers that the matrix of eigenvectors satisfies 
\begin{equation}
i\dfrac{d\hat{W}(t)}{dt}=\tilde{H}_{opt.}(t)\hat{W}(t)
\end{equation}
and the time evolution operator may be written as $\hat{U}(t,s)=\hat{W}(t)\hat{W}^{-1}(s)$,
with $\hat{W}^{-1}(t)=\hat{W}^{\dagger}(t)$. The explicit formula
for the matrix of time optimal eigenvectors is then
\begin{equation}
\hat{W}(t)=\dfrac{1}{\sqrt{2}}\left[\begin{array}{cc}
e^{2i\omega t} & -e^{2i\omega t}\\
1 & 1
\end{array}\right]
\end{equation}
Calculating identities, we can see that we have the expressions:
\begin{equation}
\hat{V}\hat{W}(t)\hat{V}=e^{-2i\omega t}\hat{W}(t)
\end{equation}
or $\hat{V}\hat{W}(t)=e^{-2i\omega t}\hat{W}(t)\hat{V}$, i.e. the
Stone-von Neumann theorem or braiding
relationship. Some other identities that exist in a similar form are:
\begin{equation}
\hat{S}^{-1}\hat{W}(t)\hat{S}^{-1}=\dfrac{1}{\sqrt{2}}\left[\begin{array}{cc}
-e^{2i\omega t} & 1\\
e^{2i\omega t} & 1
\end{array}\right]
\end{equation}
\begin{equation}
\hat{S}\hat{W}(t)\hat{S}^{-1}=\dfrac{1}{\sqrt{2}}\left[\begin{array}{cc}
e^{2i\omega t} & i\\
ie^{2i\omega t} & 1
\end{array}\right]
\end{equation}
Noting that the second identity can be modified using another unitary
operator, we choose $\hat{Z}=\left[\begin{array}{cc}
-i & 0\\
0 & 1
\end{array}\right]$, then $\hat{Z}\hat{Z}^{\dagger}=\hat{Z}^{\dagger}\hat{Z}=\mathbf{1}$,
composing this with the matrix $\hat{S}$ we have $\hat{Y}=\hat{Z}\hat{S}$,
then it is simple to show that 
\begin{equation}
\hat{Y}\hat{W}(t)\hat{Y}^{-1}=\dfrac{1}{\sqrt{2}}\left[\begin{array}{cc}
e^{2i\omega t} & 1\\
-e^{2i\omega t} & 1
\end{array}\right]=\hat{W}^{\dagger}(-t)
\end{equation}
and indeed we obtain $\hat{Y}^{2}\hat{W}(t)\hat{Y}^{-2}=\hat{W}(t)$.
In terms of the matrix operators that compose it, we have $\hat{Y}^{2}=\hat{Z}\hat{S}\hat{Z}\hat{S}=(\hat{Z}\hat{S}\hat{Z})\hat{S}$,
with $\hat{Z}\hat{S}\hat{Z}=\hat{S}^{-1}=\hat{S}^{\dagger}$, implying
that $\hat{Y}^{2}=\hat{S}^{\dagger}\hat{S}$. We can see that using
this argument it should not be too difficult to find the universal
set of gates required to implement a quantum computer using the time
optimal methodology. So far we have shown how one may use the Hadamard
matrix $\hat{V}$, the phase gate operator $\hat{Z}$ and the two
operators $\hat{S},\hat{T}$ plus their inverse to perform some very
interesting transformations between the time optimal brachistochrone
for a qubit and a number of other systems which are not solvable at
face value. Let us consider the unitary operator in the form $\hat{U}(t,s)=\hat{W}(t)\hat{W}^{\dagger}(s)=\hat{W}(t)\hat{W}^{-1}(s)$.
If we apply an isomorphic transformation on both sides, we will have:

\[
\hat{Q}\hat{U}(t,s)\hat{Q}^{-1}=\hat{Q}\hat{W}(t)\hat{Q}^{-1}\hat{Q}\hat{W}^{\dagger}(s)\hat{Q}^{-1}
\]

\begin{equation}
=\hat{W}_{Q}(t)\hat{W}_{Q}^{\dagger}(s)
\end{equation}
so we may write $\hat{U}_{Q}(t,s)=\hat{W}_{Q}(t)\hat{W}_{Q}^{\dagger}(s)$.
Let us see what results from applying this formulation to the relevant
eigenmatrices for this system. Firstly, we have:
\begin{equation}
\hat{U}_{S}(t,s)=\hat{W}_{S}(t)\hat{W}_{S}^{\dagger}(s)
\end{equation}
Note that this formula does not necessarily have $\hat{W}_{Q}^{\dagger}(s)=(\hat{W}_{Q}(s))^{\dagger}$.
Calculating, we have:
\begin{equation}
\hat{W}(t)=\dfrac{1}{\sqrt{2}}\left[\begin{array}{cc}
e^{2i\omega t} & -e^{2i\omega t}\\
1 & 1
\end{array}\right]
\end{equation}
\begin{equation}
\hat{W}_{S}(t)=\hat{S}\hat{W}(t)\hat{S}^{-1}=\dfrac{1}{\sqrt{2}}\left[\begin{array}{cc}
e^{2i\omega t} & i\\
ie^{2i\omega t} & 1
\end{array}\right]
\end{equation}
and also 

\[
\hat{W}_{S}^{\dagger}(s)=\dfrac{1}{2}\left[\begin{array}{cc}
i & -i\\
-1 & -1
\end{array}\right]\dfrac{1}{\sqrt{2}}\left[\begin{array}{cc}
e^{-2i\omega t} & 1\\
-e^{-2i\omega t} & 1
\end{array}\right]\left[\begin{array}{cc}
-i & -1\\
i & -1
\end{array}\right]
\]
\begin{equation}
=\dfrac{1}{\sqrt{2}}\left[\begin{array}{cc}
e^{-2i\omega s} & -ie^{-2i\omega s}\\
-i & 1
\end{array}\right]
\end{equation}
Note from observation we have $(\hat{W}_{S}(t))^{\dagger}=\hat{W}_{S}^{\dagger}(t)$,
which we did not assume. Computing the unitary operator we would associate
with this system, we find:

\[
\hat{U}_{S}(t,s)=\hat{W}_{S}(t)\hat{W}_{S}^{\dagger}(s)
\]

\[
=\dfrac{1}{2}\left[\begin{array}{cc}
e^{2i\omega t} & i\\
ie^{2i\omega t} & 1
\end{array}\right]\left[\begin{array}{cc}
e^{-2i\omega s} & -ie^{-2i\omega s}\\
-i & 1
\end{array}\right]
\]

\begin{equation}
=\dfrac{1}{2}\left[\begin{array}{cc}
e^{2i\omega(t-s)}+1 & -i(e^{2i\omega(t-s)}-1)\\
i(e^{2i\omega(t-s)}-1) & e^{2i\omega(t-s)}+1
\end{array}\right]
\end{equation}
which we rewrite as:
\begin{equation}
\hat{U}_{S}(\phi)=e^{i\phi}\left[\begin{array}{cc}
\cos\phi & \sin\phi\\
-\sin\phi & \cos\phi
\end{array}\right]
\end{equation}
which agrees with results from the earlier calculation. We can see
how the eigenmatrix is rotated into the new reference frame to generate
the unitary operator; the same method may be extended to the other
matrices we have considered in this paper, being $\hat{T},\hat{V},\hat{Z}$
and all combinations of inverses, concatenations of products etc.
Let us now examine some elementary results and offer some comments
on why this method is so effective compared to some others. We shall
then discuss how this method may be extended to hyperbolic geometries
and how this impacts on the calculation of Vilenkin \cite{vilenkin1978special}. 

\section{Hyperbolic Equivalent of the Brachistochrone}

Let us now talk about how this method may be extended to hyperbolic
space. As is well known, SU(2) is associated to a spherical geometry;
we are interested in how the SU(2)$\leftrightarrow$SU(1,1) equivalence
may be exploited. In general terms, we expect that it will take the
form of a rotation into imaginary time, in the same way that a Wick
rotation \cite{wick1954properties} allows us to take a complex path integral
for a quantum space and use this to generate a real path integral
for a stochastic space, in imaginary time. Writing now $i\left|\dot{\Psi}\right\rangle =i\dfrac{d}{dt}\left|\Psi\right\rangle =\tilde{H}(t)\left|\Psi\right\rangle $,
it is simple to see that we will have
\begin{equation}
\int1dt=\int\dfrac{\sqrt{\left\langle \Psi\right|\left.\tilde{H}(1-\hat{P})\tilde{H}\right.\left|\Psi\right\rangle }}{\Delta E}
\end{equation}
or, upon using the standard expression for the projective operator,
i.e. $\hat{P}=\left|\Psi\right\rangle \left\langle \Psi\right|$,
the energy dispersion is then
\begin{equation}
\Delta E=\sqrt{\left\langle \Psi\right|\left.\tilde{H}^{2}(t)\left|\Psi\right\rangle -\left(\left\langle \Psi\right|\tilde{H}(t)\left|\Psi\right\rangle \right)^{2}\right.}
\end{equation}
which we recognise as the standard expression for the variance in
the energy as represented in the quantum system. Let us now examine
the nature of complexification with respect to this system. We have
a quantum state which is dependent on time via $\left|\Psi(t)\right\rangle $,
we have also variously a Hamiltonian matrix $\tilde{H}(t)$ and a
constraint $\tilde{F}(t)$, all of these objects are required in order
to specify the state of the system. The quantum brachistochrone, which
is equivalent to the von Neumann equation for the operator $\tilde{H}(t)+\tilde{F}(t)$,
may be written as \cite{carlini2005quantum, morrison2008time} 
\begin{equation}
i\dfrac{d}{dt}(\tilde{H}(t)+\tilde{F}(t))=\tilde{H}(t)\tilde{F}(t)-\tilde{F}(t)\tilde{H}(t)
\end{equation}
and the unitary operator of the system is defined by $\hat{U}(t,s)=\exp\left(-i\int_{s}^{t}\tilde{H}(\tau)d\tau\right)$.
We are in the relatively comfortable situation of knowing (roughly!)
what the answers will be. Let us borrow from Vilenkin \cite{vilenkin1978special}
and examine the nature of the unitary matrices used in their analysis
of SU(1,1). Following earlier analysis, we have e.g.
\begin{equation}
\hat{u}_{2}=\left[\begin{array}{cc}
\cosh\dfrac{t}{2} & i\sinh\dfrac{t}{2}\\
\\
-i\sinh\dfrac{t}{2} & \cosh\dfrac{t}{2}
\end{array}\right]
\end{equation}
as a basic unitary operator in SU(1,1), whereas the previous section
has shown that matrices of the form
\begin{equation}
\hat{U}_{S}(t,s)=e^{i\phi}\left[\begin{array}{cc}
\cos\phi & \sin\phi\\
-\sin\phi & \cos\phi
\end{array}\right]
\end{equation}
are the relevant concern for SU(2), with $\phi=\omega(t-s)$. Let
us examine the relation between these two groups of matrices. Obviously
we must look at extensions into complex time via $t\rightarrow it$.
If we neglect the global phase in $\hat{U}_{S}(t,s)$, we can see
that 

\[
\hat{U}_{S}(it,0)=e^{-\omega t}\left[\begin{array}{cc}
\cos i\omega t & \sin i\omega t\\
-\sin i\omega t & \cos i\omega t
\end{array}\right]
\]

\begin{equation}
=e^{-\omega t}\left[\begin{array}{cc}
\cosh\omega t & i\sinh\omega t\\
-i\sinh\omega t & \cosh\omega t
\end{array}\right]
\end{equation}
which, up to scaling by the factor $e^{-\omega t}$, is equivalent
to the matrix $\hat{u}_{2}$ used in Vilenkin \cite{vilenkin1978special}.
We can see the nature of the transformation between spherical and
hyperbolic spaces in this formula. Let us now consider further the
implications for the action principle. Indeed, the quantum brachistochrone,
by its nature, implies optimality in terms of time taken between any
two points on the complex projective space. It seems sensible to then
conjecture that, in order to obtain the hyperbolic counterparts of
the spherical operators, we must look at optimality in terms of imaginary
time. Note that the operators that we obtain, such as the operator
$\hat{U}_{S}(it,0)$, are not unitary in the normal sense. We shall
comment on this. Note that if we write $\hat{U}_{S}(it,is)=\hat{\mathcal{U}}_{S}(t,s)$,
we have time translation invariance in the form:

\[
\hat{\mathcal{U}}_{S}(t,s)=\hat{U}_{S}(it,is)=\hat{U}_{S}(i(t-s),0)
\]

\begin{equation}
=e^{-\omega(t-s)}\left[\begin{array}{cc}
\cosh[\omega(t-s)] & i\sinh[\omega(t-s)]\\
-i\sinh[\omega(t-s)] & \cosh[\omega(t-s)]
\end{array}\right]
\end{equation}
Expanding by using the double angle formulae for hyperbolic sines
and cosines, we find:

\begin{equation}
\hat{\mathcal{U}}_{S}(t,s)=\hat{\mathcal{U}}_{S}(t,0)\hat{\mathcal{U}}_{S}(0,s)
\end{equation}
with
\begin{equation}
\hat{\mathcal{U}}_{S}(t,0)=e^{-\omega t}\left[\begin{array}{cc}
\cosh\omega t & i\sinh\omega t\\
-i\sinh\omega t & \cosh\omega t
\end{array}\right]
\end{equation}
Obviously we have $\hat{\mathcal{U}}_{S}(0,s)=\hat{\mathcal{U}}_{S}(-s,0)$
which means that this space is not a Hermitian space. Let us now examine
the nature of the time evolution equation in complex time and see
what can be learnt. Firstly, we know that the time evolution operator
in the normal space may be written in exponential form as:

\begin{equation}
\hat{U}_{S}(t,s)=\exp\left(-i\int_{s}^{t}\tilde{H}_{S}(\tau)d\tau\right)=\hat{W}_{S}(t)\hat{W}_{S}^{-1}(s)
\end{equation}
with fundamental eigenmatrix:
\begin{equation}
\hat{W}_{S}(t)=\dfrac{1}{\sqrt{2}}\left[\begin{array}{cc}
e^{2i\omega t} & i\\
ie^{2i\omega t} & 1
\end{array}\right]
\end{equation}
and Hamiltonian operator defined by

\begin{equation}
\tilde{H}_{S}(t)=R\left[\begin{array}{cc}
-\cos(2\omega t) & \sin(2\omega t)\\
\sin(2\omega t) & \cos(2\omega t)
\end{array}\right]
\end{equation}
also $\hat{W}_{S}^{-1}(s)=\hat{W}_{S}^{\dagger}(t)$. Let us see what
happens if we perform the change into complex time. The evolution
operator will be altered in the following way:

\[
\hat{\mathcal{U}}_{S}(t,s)=\hat{U}_{S}(it,is)=\exp\left(-i\int_{is}^{it}\tilde{H}_{S}(\tau)d\tau\right)
\]

\begin{equation}
=\exp\left(-\int_{s}^{t}\tilde{H}_{S}(i\tau)d(i\tau)\right)
\end{equation}
We would hope that $\hat{U}_{S}(it,is)=\hat{W}_{S}(it)\hat{W}_{S}^{\dagger}(-is)$.
Let us see if this is indeed the case. Calculating the right-hand
side, we have immediately:
\begin{equation}
\hat{W}_{S}(it)\hat{W}_{S}^{\dagger}(-is)=e^{-\phi}\left[\begin{array}{cc}
\cosh\phi & i\sinh\phi\\
-i\sinh\phi & \cosh\phi
\end{array}\right]
\end{equation}
which we notice is the same transformation we previously calculated,
with $\phi=\omega(t-s)$. The technique works effectively as can be
seen. In terms of the Hamiltonian operator on the hyperbolic space,
we have

\[
\tilde{\mathcal{H}}_{S}(t)=\tilde{H}_{S}(it)=R\left[\begin{array}{cc}
-\cos(2\omega it) & \sin(2\omega it)\\
\sin(2\omega it) & \cos(2\omega it)
\end{array}\right]
\]

\begin{equation}
=R\left[\begin{array}{cc}
-\cosh(2\omega t) & i\sinh(2\omega t)\\
i\sinh(2\omega t) & \cosh(2\omega t)
\end{array}\right]
\end{equation}
Let us now consider the effect of this type of transformation on the
quantum brachistochrone equations. This set of coupled DEs is described
by the system
\begin{equation}
i\dfrac{d}{dt}(\tilde{H}(t)+\tilde{F}(t))=[\tilde{H}(t),\tilde{F}(t)]
\end{equation}
Let us examine the nature of differentiating e.g. $\dfrac{d}{dt}(\tilde{H}(it)+\tilde{F}(it))$,
the left-hand side of the above equation will then be changed to:
\begin{equation}
-\dfrac{d}{dt}(\tilde{H}(it)+\tilde{F}(it))=[\tilde{H}(it),\tilde{F}(it)]
\end{equation}
We may further transform this equation by a constant matrix $\hat{S}$,
and we would hope to find the set of coupled differential equations
defined by:
\begin{equation}
-\dfrac{d}{dt}(\tilde{\mathcal{H}}_{S}(t)+\tilde{\mathcal{F}}_{S}(t))=[\tilde{\mathcal{H}}_{S}(t),\tilde{\mathcal{F}}_{S}(t)]
\end{equation}
Let us check that this is indeed the case once more. For our matrix
$\tilde{\mathcal{H}}_{S}(t)$, we have associated constraint law $\tilde{\mathcal{F}}_{S}(t)=\Omega\hat{\sigma}_{y}$,
obviously these two operators satisfy $\mathrm{Tr}(\tilde{\mathcal{H}}_{S}(t)\tilde{\mathcal{F}}_{S}(t))=0$,
so this is at least working. We have $\tilde{\mathcal{F}}_{S}(t)=\tilde{\mathcal{F}}_{0}$,
so we would hope that we are able to satisfy
\begin{equation}
-\dfrac{d\tilde{\mathcal{H}}_{S}(t)}{dt}=[\tilde{\mathcal{H}}_{S}(t),\tilde{\mathcal{F}}_{0}]
\end{equation}
Calculating the left-hand side, we have 
\begin{equation}
-\dfrac{d\tilde{\mathcal{H}}_{S}(t)}{dt}=-2\omega R\left[\begin{array}{cc}
-\sinh(2\omega t) & i\cosh(2\omega t)\\
i\cosh(2\omega t) & \sinh(2\omega t)
\end{array}\right]
\end{equation}
whereas the right-hand side of this expression reads as:
\begin{equation}
[\tilde{\mathcal{H}}_{S}(t),\tilde{\mathcal{F}}_{0}]=2\Omega R\left[\begin{array}{cc}
-\sinh(2\omega t) & i\cosh(2\omega t)\\
i\cosh(2\omega t) & \sinh(2\omega t)
\end{array}\right]
\end{equation}
hence we have a solution for $\Omega=-\omega$. This is the appropriate
method to use to calculate a brachistochrone in a hyperbolic space.
Let us now examine the nature of the isotropic constraint, which in
the standard quantum setting reads as:
\begin{equation}
\mathrm{Tr}\left(\dfrac{\tilde{H}^{2}}{2}\right)=k<\infty
\end{equation}
If we take the formula we derived above, we have 
\begin{equation}
\tilde{\mathcal{H}}_{S}(t)=\tilde{H}_{S}(it)=R\left[\begin{array}{cc}
-\cosh(2\omega t) & i\sinh(2\omega t)\\
i\sinh(2\omega t) & \cosh(2\omega t)
\end{array}\right]
\end{equation}
Naiively computing the isotropic constraint, we obtain:
\begin{equation}
\mathrm{Tr}\left(\dfrac{\tilde{H}^{2}}{2}\right)=R^{2}[\cosh^{2}(2\omega t)-\sinh^{2}(2\omega t)]=R^{2}<\infty
\end{equation}
We can see that this method will be effective for finding brachistochrones
on these types of quasi-unitary spaces. Finally, we shall comment
on the nature of the action principle in a hyperbolic space. Following
Carlini et. al \cite{carlini2006time}, for a standard time optimal quantum
system, the energy dispersion is written as:
\begin{equation}
\Delta E=\sqrt{\left\langle \Psi\right|\left.\tilde{H}^{2}(t)\left|\Psi\right\rangle -\left(\left\langle \Psi\right|\tilde{H}(t)\left|\Psi\right\rangle \right)^{2}\right.}
\end{equation}
We have already shown that the trace conditions and isotropic relation
are essentially unchanged by moving to complex time. We are posed,
therefore, with the question of what exactly constitutes the state
vector in this type of hyperbolic space.

\section{Transformed Hamiltonian Matrix}

Let us calculate the effect of implementing the transformation

\begin{equation}
\hat{W}_{S}(t)=\dfrac{1}{\sqrt{2}}\left[\begin{array}{cc}
e^{2i\omega t} & i\\
ie^{2i\omega t} & 1
\end{array}\right]
\end{equation}
Calculating $\hat{W}_{S}^{-1}(t)\tilde{H}_{S}(t)\hat{W}_{S}(t)$ for
the standard SU(2) system, we have 
\begin{equation}
\hat{W}_{S}^{-1}(t)\tilde{H}_{S}(t)\hat{W}_{S}(t)=R\hat{\sigma}_{y}
\end{equation}
We can see that this transformation takes the Hamiltonian operator
into a reference frame where the matrix is a constant. However, this
is not the matrix of eigenvalues with respect to the angular momentum.
In this case, exponentiating the operator gives
\begin{equation}
\exp\left(-itR\hat{\sigma}_{y}\right)=\left[\begin{array}{cc}
\cos Rt & -\sin Rt\\
\sin Rt & \cos Rt
\end{array}\right]
\end{equation}
The unitary operator for this system is given by:
\begin{equation}
\hat{U}_{S}(t,s)=e^{i\phi}\left[\begin{array}{cc}
\cos\phi & \sin\phi\\
-\sin\phi & \cos\phi
\end{array}\right]
\end{equation}
We can see that there is an implicit connection shown through this
transformation. Obviously we could go through all the different transformations
we have examined and examine the nature of the transformation laws.
It is interesting that in this case we are transforming to a constant
matrix. For these types of small systems the essential degeneracy
of the system means that there is overlap between some of the objects,
which results in the pleasing amount of symmetries we have generated
throughout this paper. We have not made an effort to make this an
exhaustive proof, as it is still unclear how some of the objects relate
to one another, but at least the major part of the calculation has
cleared up these issues which became evident through other papers.

\section{Adjoint Representations}

In this section we will examine the nature of the adjoint representations
generated by the time optimal unitary operators we have calculated
in this paper. The basic identity we shall use to compute this will
be of the form:

\begin{equation}
\hat{U}_{Q}(\phi)\vec{\mathbf{\sigma}}\hat{U}_{Q}^{\dagger}(\phi)=\hat{R}_{Q}(\phi)\vec{\mathbf{\sigma}}
\end{equation}
In this formula $Q$ represents the transformation with which we are
transforming the time optimal unitary with respect to, i.e. $\hat{S},\hat{T},\hat{V}$;
$\vec{\mathbf{\sigma}}$ is a vector of Pauli matrices, or a basic
spinor; $\hat{U}_{Q}(\phi)$ is the time optimal unitary transformed
by the operator $\hat{Q}$; and $\hat{R}_{Q}(\phi)$ is a matrix which
we call 'adjoint', which gives the transformation of the spinor in
a linear form. We shall illustrate how this formula is applied through
a simple example then list some results. Let us take the time optimal
unitary as our prototypical example. In this case it is simple to
show that we have the vector-matrix equation
\begin{equation}
\hat{U}(\phi)\vec{\mathbf{\sigma}}\hat{U}^{\dagger}(\phi)=\left[\begin{array}{ccc}
\cos2\phi & -\sin2\phi & 0\\
\sin2\phi & \cos2\phi & 0\\
0 & 0 & 1
\end{array}\right]\left[\begin{array}{c}
\hat{\sigma}_{x}\\
\hat{\sigma}_{y}\\
\hat{\sigma}_{z}
\end{array}\right]
\end{equation}
We
can see that, in terms of the adjoint representation, we have a simple
connection between the groups we have been examining on SU(2) and
the equivalent matrices in SO(3) that represent rotation in 3-D space.

\section{Conclusions}

We have shown in this paper how one may use concepts from time optimal
control to derive a large number of results relating to the matrix
groups that exist on the associated space. Indeed, we have produced sufficient logic gates in order to discuss higher order concepts of computation. It is well-known in the field of quantum information theory that a set of such gates is universal for computation. In our case, we are able to reduce computation to a single operator, and unitary transforms actuated through eigenvalue decompositions. Using the quantum brachistochrone principle we have been able to easily identify the symmetries of the system, and identify the dynamics. 

Hillis discussed the Connection Machine and his work with Feynman in \cite{hillis1985connection}. The Connection Machine was an early type of multiparallel processor, based on an elegant and complex architecture. Amongst other things, Feynman adapted this computer to solve questions in neural networks and quantum chromodynamics. Feynman was allocated one task- something that didn't sound like 'baloney', and that turned out to be analysis of the router in a 20-dimensional hypercube. Willis states that "By the end of...1983, Richard (Feynman) had completed his analysis of the behavior of the router, and much to our surprise and amusement, he presented his answer in the form of a set of partial differential equations. To a physicist this may seem natural, but to a computer designer, treating a set of boolean circuits as a continuous, differentiable system is a bit strange. Feynman’s router equations were in terms of variables representing continuous quantities such as “the average number of 1 bits in a message address.” I was much more accustomed to seeing analysis in terms of inductive proof and case analysis than taking the derivative of “the number of 1’s” with respect to time.". In a sense, the theory of time optimal control is very much in the spirit of Feynman's initial investigations into computing, whether through the development of machines at Los Alamos, through to the work with the Connection Machine. In our machine, we are primarily concerned with such concepts, with the caveat that, instead of looking at averages, as by the Ehrenfest theorem and in Feynman's analysis of the router design, we are looking at the rates of change of data streams directly through use of quantum mechanics. This is given by the von Neumann perspective of quantum mechanics, which in our case takes the form of the quantum brachistochrone equation. 

Quantum information science is faced with a number of engineering difficulties in terms of the digital processing of data, especially as we head to more and more regimes where quantum mechanics is expected to dominate. Indeed, what can be seen is an aspect of the "war of the currents", in microcosm. Time optimal quantum control offers an easy option out. The continuation of previously discarded ideas such as analog processing and computation, photonic circuits, and AC computing in general becomes a more appealing prospect the closer we get to the nanoscale of atomic fabrication. By showing that we can use a clever arrangement of time optimal unitary or pseudounitary operators to recover logic gates, we hope that the general utility of this perspective of quantum mechanics can be brought to light. Much of the work on analog computation, differential integrators and the like was laid to rest post World War II, with the advent of the age of digital computation. Our method harks back to this, and we must make it clear that this is a very different form of data processing that we are currently familiar with, both in the quantum and classical realm. More generally, it might be hoped that by analysis of the quantum brachistochrone problem for processors, one might be able to develop a greater theory of time dependent data flow. We must start with a single qubit in order to achieve such grand aims. In that we have surely been successful.

In this case we have stuck
to the simplest of examples, i.e. qubits and their hyperbolic equivalents;
however, this is nothing more than a convenient sandbox where we may
apply our concepts. The methods we have shown are generally applicable,
and it is hoped that this paper provokes more effort in the direction
of understanding the nature of the connections between different groups
of matrices for small spaces. As can be seen, there is a coherent
picture provided by the quantum brachistochrone and time optimal control
theory. The results we have demonstrated, while applicable only to
the problems we have considered, indicate the outlines of a more general
theory which shall be explored in further papers. In particular, the
space of SU(2) and hyperbolic equivalent SU(1,1) are a little too
simplistic to be descriptive of reality. One can see this in moving
e.g. to SU(3) \cite{morrison2019time2}, where the number of optimal transformations
is increased, and we no longer have a single brachistochrone that
takes the state to a diagonal reference frame. Regardless of this
essential complication, we have succeeded in demonstrating that the
quantum brachistochrone equation is a concise, succinct and elegant
approach towards understanding these types of finite quantum groups.

\section{Acknowlegements}
This project was supported under ARC Research Excellence Scholarships at the University of Technology, Sydney. The author acknowledges the support and assistance of Dr. Mark Craddock and Prof. Anthony Dooley.

\section{Data Sharing Declaration}
The author declares that data sharing is not applicable to this article as no datasets were generated or analysed during the current study.

\section{Conflicts of Interest}
The author has no conflicts of interest to declare that are relevant to the content of this article.

\bibliography{sn-bibliography}

\end{document}